\documentclass{article}
\usepackage{graphicx}
\usepackage[T1]{fontenc}
\usepackage{amsmath}
\usepackage{graphicx}
\usepackage{txfonts}
\usepackage{natbib} % bibliographie, avec \citet et \citep
\bibpunct{(}{)}{;}{a}{}{,} % to follow the A&A style

\newcommand{\be}{\begin{equation}}
\newcommand{\ee}{\end{equation}}

\newcommand{\lapa}{\nabla^2_{\theta,\phi}}
\begin{document}

\title{General solution for the vacuum electromagnetic field in the
  surroundings of a rotating star. 
   %({\bf 2013_psr_mag_v3.tex})
   }

   \author{S. Bonazzola and  F. Mottez  \\ LUTH, Observatoire de Paris-psl, \\CNRS, Universit\'e Paris Diderot,
   	5 place Jules Janssen, 92190 Meudon, France.\\
          \and
          J. Heyvaerts \\ Observatoire Astronomique, Universit\'e de Strasbourg,
         \\ 11, rue de l'Universit\'e, 67000 Strasbourg, France.
          }

%   \institute{LUTH, Observatoire de Paris-psl, CNRS, Universit\'e Paris Diderot,
%              5 place Jules Janssen, 92190 Meudon, France.\\
%              \email{fabrice.mottez@obspm.fr}
%         \and
%             Observatoire Astronomique, Universit\'e de Strasbourg,
%		11, rue de l'Universit\'e, 67000 Strasbourg, France.\\
%             \email{jean.heyvaerts@astro.unistra.fr}}

   %Version du~: \today          
   %\date{Received September 15, 1996; accepted March 16, 1997}

%\abstract{}{}{}{}{} 
% 5 {} token are mandatory
 
  %\abstract{} 
  %% context heading (optional)
  %% {} leave it empty if necessary  
  % {}%heading}
  %% aims heading (mandatory)
  \maketitle
  
  \section{Abstract}{Many recent observations of pulsars and magnetars can be interpreted in terms of neutron stars (NS) with multipole electromagnetic fields. As a first approximation, we investigate the  multipole magnetic and electric fields in the environment 
  of a rotating star when this environment is deprived of plasma.}
  %% methods heading (mandatory)
  {We compute a  multipole expansion of the electromagnetic field in vacuum for a given magnetic field on the conducting surface of the rotating star. Then, we consider a few consequences of multipole fields of pulsars.}
  %% results heading (mandatory)
  {  We provide an explicit form of the solution. For each spherical
  harmonic of the magnetic field,  the expansion contains a finite  number of terms.  A multipole magnetic field can provide an explanation for the stable sub-structures of pulses, and they offer a solution to the problem of current closure in pulsar magnetospheres.
  }
  % {}%conclusions
  {This computation generalises the widely used model of a rotating star in vacuum with a
  dipole field. It can be especially useful as a first approximation
  to the electromagnetic environment of a compact star, for instance a
  neutron star, with an arbitrarily magnetic field. }
%{\it Applications}
%   \keywords{pulsars -- exoplanets-- magnetosphere }
%   \titlerunning{magnetic thrust}
%   \authorrunning{Mottez and Heyvaerts}
%   \maketitle

\section{Introduction}

Dipole magnetic fields have two important properties that contribute to their success in the modelling of a pulsar magnetosphere: 
dipole fields dominate higher order multipole fields at large distances from the neutron star, and they are computationally simpler.
Mostly based on the consideration of spin-up lines in the $P-\dot{P}$ diagram, \cite{Arons_1993} showed that low-altitude magnetic fields 
of pulsars are dominated by their dipole component, the non-dipole component not exceeding 40\% of the dipole field.
However, several observations tend to show that multipole magnetic fields cannot be neglected in every pulsar.

\cite{Gotthelf_2013}
 measured period derivatives for the pulsar PSR J0821-4300 . It is a central compact object (CCO) in a supernova remnant. They  found exceptionally weak dipole magnetic field components for a young neutron star, about $10^{10}$ G. Antipodal
surface hot spots with different temperatures and areas were deduced from the X-ray spectrum and pulse profiles.
Such non-uniform surface temperature appears to require strong crustal magnetic fields, probably
toroidal or quadrupolar components much stronger than the external dipole.

The pulsar J2144-3933,  with a period of 8.51 s,  is beyond the death-line in the $P-B_s$ diagram and according to standard emission models. 
The pulsar should not emit radio-waves. Death-line models strongly depend on magnetic field lines curvatures. For a given surface
magnetic dipole field strength, pulsars with strong multipolar field
components have a highly curved field near the stellar surface that might
permit the radio emission of the pulsar J2144-3933 \citep{Young_1999}. 
Indeed, \cite{Harding_2011} has shown that a simply offset dipole field can increase the pair-cascade efficiency, and lower the death-line
in the $P-\dot{P}$ diagram.

Multi-wavelength observations of  intermittent 
radio emissions from rotation-powered pulsars beyond the pair-cascade death 
line, of the pulse profile of the magnetar SGR 1900+14 after its 1998 August 27 
giant flare and  of the X-ray spectral features of PSR J0821-4300 and SGR 0418+5729, 
suggest that the magnetic fields of non-accreting neutron stars are not purely 
dipolar and may contain higher order multipoles \citep{Mastrano_2013}.

\cite{Guver_2011}  analysed upper bound on the spin-down rate and the high signal-to-noise ratio XMM-Newton spectra of the soft gamma-ray repeater SGR 0418+5729. They found a low surface magnetic field  in comparison to other magnetars: $10^{14}$ G. In connection to the spin-down limits, this implies a significantly multipole structure of the magnetic field.

\begin{figure}
%\resizebox{\hsize}{!}{\includegraphics{fig_ailealffig1bis.eps}}
\resizebox{\hsize}{!}{\includegraphics[angle=0]{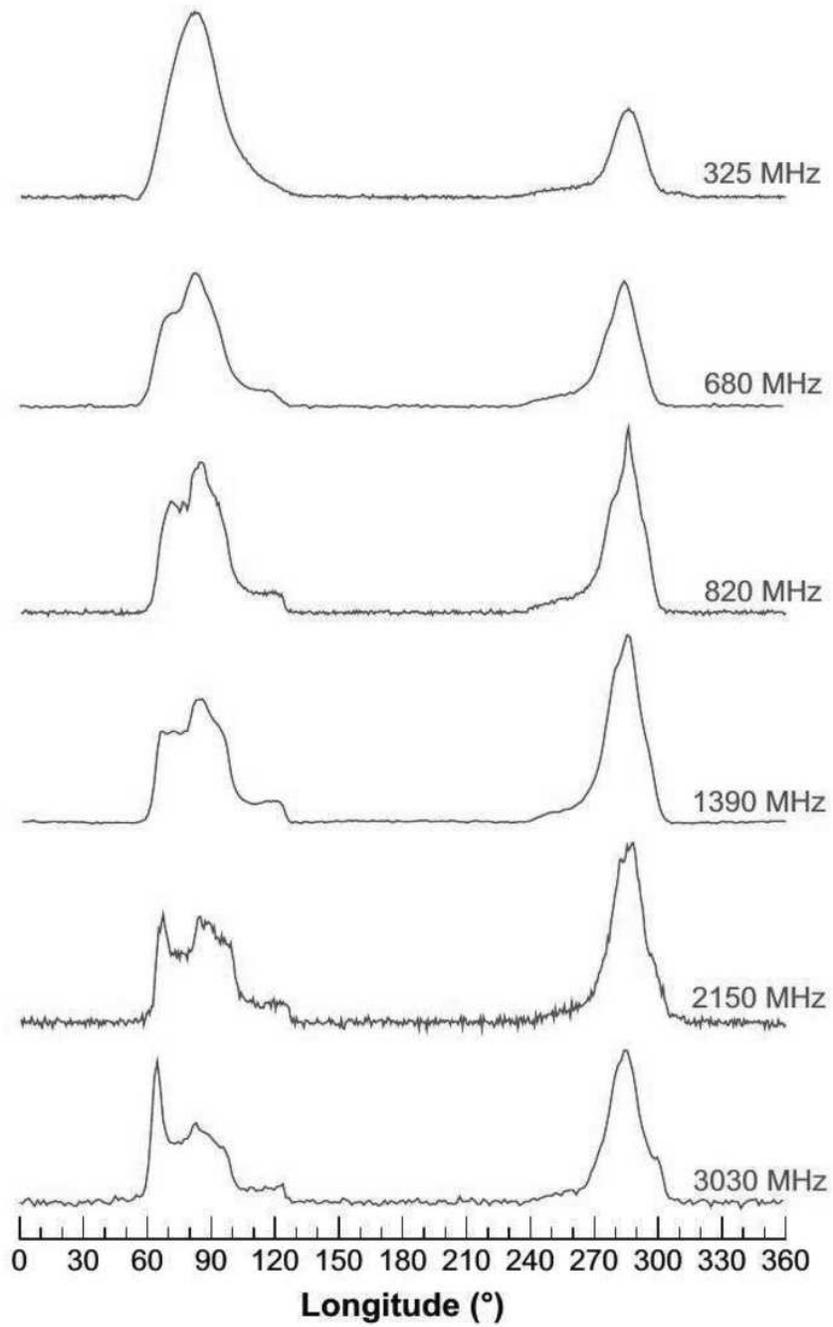}} %{fig_psr_binaire_A_radio_Kramer_2006.ps}}
\caption{Pulse profiles of PSR J0737-3039A at various radio frequencies. From \citep{Kramer_2008_double_psr}.}\label{fig_psr_binaire_A_radio_Kramer_2006}
\end{figure}

Most attempts to model the pulses profiles of pulsars are based on dipole magnetic fields. But some features of these profiles resist the models. 
Let us consider, for instance, the brightest pulsar A of the two pulsars binary system PSR J0737-3039. The radio pulse profiles of PSR J0737-3039A consist of two peaks shown in Fig. \ref{fig_psr_binaire_A_radio_Kramer_2006} \citep{Kramer_2008_double_psr}. Geometrical models have been produced with best-fit  one and two pole models  \citep{Ferdman_2013}, two Poles Caustics (TPC), Outer Gap (OG)   \citep{Guillemot2013_psr_double_fermi} and a retarded  vacuum dipole polar cap \citep{Perera_2014}.  With these models, one can reconstruct the main angles defining the orbital plane, rotation axis and magnetic inclination of the pulsar, as well as the general shape of the pulses. For instance \cite{Ferdman_2013} could reconstruct a Gaussian fit, and \cite{Perera_2014} considered pulse width at four intensity levels. All these models involve a dipole magnetic field. They found that the two peaks are more likely to be associated with the two poles. But the peaks (especially the less intense one) show sub-structures that do not enter into their models. The sub-structures (a spiky plateau above the 75\% intensity level before the main maximum, and a plateau at the 10 \% level after the main maximum) occupy a significant proportion of the total phase angle. It is quite possible that these sub-structures are associated with multipole components of the electromagnetic environment of the neutron star.  

\begin{figure}
%\resizebox{\hsize}{!}{\includegraphics{fig_ailealffig1bis.eps}}
\resizebox{\hsize}{!}{\includegraphics[angle=0]{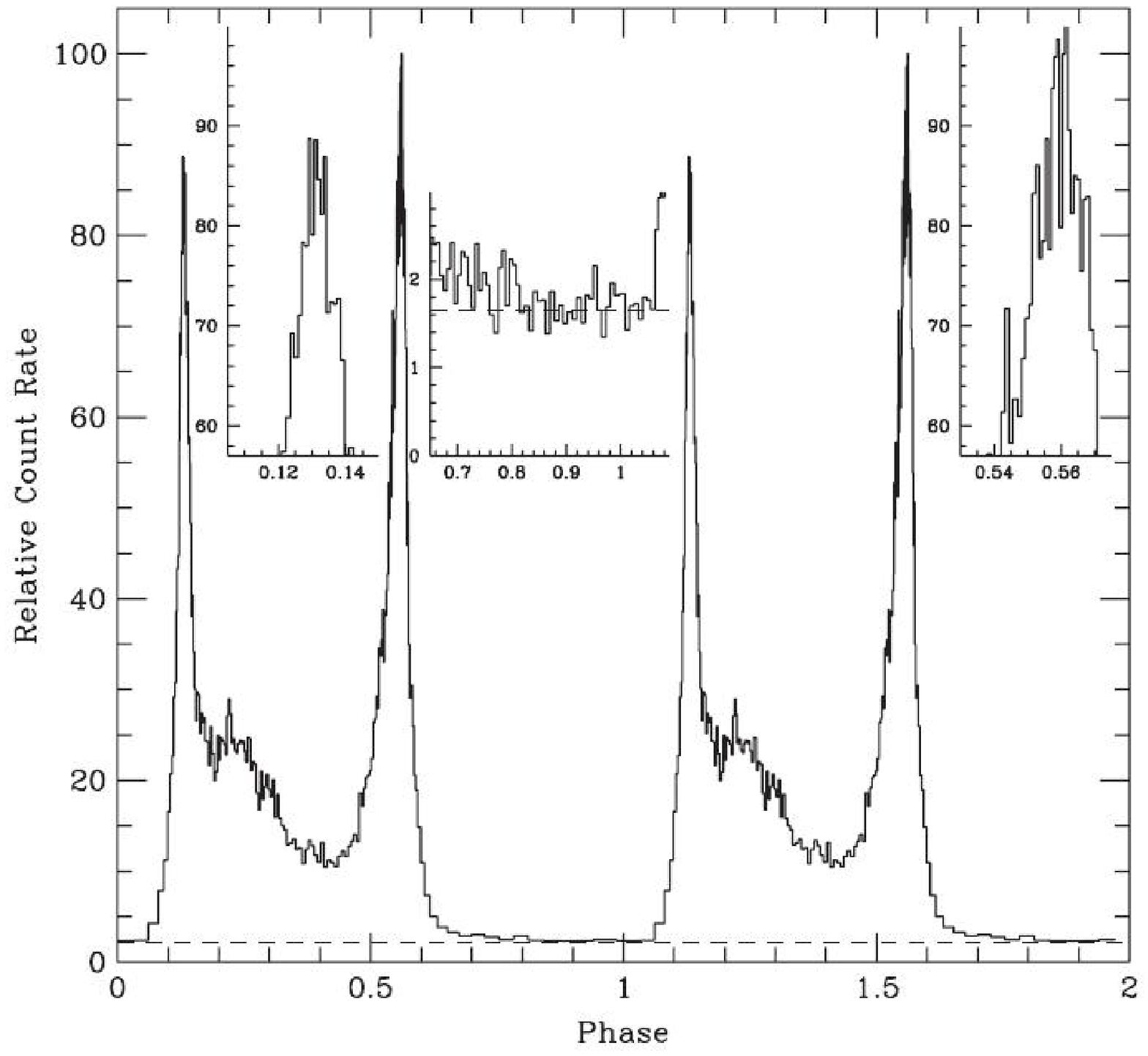}} %{fig_pulse_vela_fermi_2009.ps}}
\caption{Vela broadband (E = 0.1 - 10 GeV) pulse profile. Two pulse periods are shown. The dashed line shows the background
level, as estimated from a surrounding annulus during the off-pulse phase. Insets
show the pulse shape near the peaks and in the off-pulse region
  (from Abdo et al. 2009 %\citet{Abdo_2009_vela_fermi}
  ).}\label{fig_pulse_vela_fermi_2009}
\end{figure}

Another example can be seen in the gamma rays profile of the Vela pulsar revealed by the Fermi-LAT telescope \citep{Abdo_2009_vela_fermi} displayed in Fig. \ref{fig_pulse_vela_fermi_2009}. Again, this profile contains large sub-structures. It also exhibits shorter sub-structures, which are visible in the enlarged insets.
Here again, multipole components might be a cause of pulses sub-structures. 

As recalled by  \cite{Perera_2014}, in general, pulsar magnetosphere models are constructed at the following two
limits: (a) a vacuum limit \citep{Deutsch_1955}, and (b) a force-free magnetohydrodynamics (MHD) limit
with a plasma-filled magnetosphere \citep{Spitkovsky_2006}. However, a true magnetosphere operates
between these two limits.  We could expect that the MHD solutions are more realistic, but \cite{Harding_2011} found that the rotating dipole magnetosphere in vacuum, in many cases, provides better fits to observed gamma rays pulse profiles than the force-free magnetosphere. This is for instance what they found for Vela. 
This shows that the vacuum magnetosphere is still a useful approximation in pulsar physics. 

Considering this general remark and the possible relevance of multipole electromagnetic fields to pulsar models, we present an analytically exact model of the 
vacuum magnetosphere where the neutron star magnetic field is expanded in multipole components. 
  
  Suitable boundary conditions are taken into account for an
  oblique rotator with a conducting surface. This algorithm allows us to describe
  electromagnetic fields with $l,m$ quantum numbers as high as $100$
($ l \ge m $). This algorithm is a generalisation of the one described by
  \cite{Deutsch_1955} for a simple magnetic dipole ($l=1$). 

 In section \ref{sec_methods}, we present the method of resolution of the Maxwell equations
 and their boundary conditions. 
In the section \ref{sec_axial}, we present the parallel solutions ($m=0$) for any value
  of $l$.
Section \ref{sec_non_axial} contains the general solution 
of the  Maxwell equation with the required boundary conditions
for given quantum numbers  $l,m \; m \ge 1$. 
The numbers $m>0$ correspond to the perpendicular case, i.e. where the axis of the mutipole is orthogonal to the axis of the neutron
  star.
Thanks to the linearity of the Maxwell equation, the general solution is
 a linear combination of the perpendicular and parallel solutions.
 The matching conditions are applied in section \ref{sec_matching_conditions}.
Details of the analytical calculations are presented in appendices \ref{sec_solution_Etheta_nul}-\ref{matching_condition_is_ok}.

After this derivation, two applications of multipoles are suggested. The first concerns the problem of the pulsar current closure, and the second 
concerns the pulse profile of pulsars such as PSR J0737-3039A and Vela. 

\section{Methods} \label{sec_methods}

Vectors are expanded in spherical coordinates of axis $z$ parallel to the angular velocity vector ${\bf \Omega}$
of the neutron star.

Following \cite{Bonazzola_2007} let us define the components of the magnetic field as 
 the usual radial component $B_r$, and  
two scalar fields, $\eta$ and $\mu$, such that 
\begin{eqnarray} \label{def_eta_mu}
B_\theta &=& \partial_\theta \eta -\frac{1}{\sin \theta} \partial_\phi \mu, \nonumber \\
B_\phi &=& +\frac{1}{\sin \theta} \partial_\phi \eta  +\partial_\theta \mu.
\end{eqnarray}
The magnetic fields, and $\mu$ and $\eta$, are also related through the relations
\begin{eqnarray} \label{magique}
\nabla_{\theta \phi}^2 \eta &=& \partial_\theta B_\theta +\frac{\cos \theta}{\sin \theta}B_\theta +\frac{1}{\sin \theta} \partial_\phi B_\phi, \nonumber \\
\nabla_{\theta \phi}^2 \mu &=& \partial_\theta B_\phi +\frac{\cos \theta}{\sin \theta} B_\phi -\frac{1}{\sin \theta} \partial_\phi B_\theta \nonumber \\ 
&=& \frac{1}{\sin \theta} [\partial_\theta (B_\phi \sin \theta) - \partial_\phi B_\theta].
\end{eqnarray}
The  magnetic field associated with $\mu$, noted $B^{TM}$ is transverse/toroidal (i.e. $B_r^{TM}=0$)
as well as the electric derived from $\eta$, noted $E^{TE}$ (i.e. $E_r^{TE}=0$), defining the poloidal electromagnetic field. The magnetic and electric field ${\bf B}$, $ {\bf E} $ around a magnetised spinning star ${\mathbf{B}} $, can be considered the
  sum of the poloidal field (${\mathbf{B}}^{TE} $, ${\mathbf{E}}^{TE}$) and a toroidal field
(${\mathbf{B}}^{TM}$, ${\mathbf{E}}^{TM}$ ),
\begin{equation}
{\mathbf{B}} = {\mathbf{B}}^{TE} + {\mathbf{B}}^{TM}, \qquad \qquad \qquad
{\mathbf{E}} = {\mathbf{E}}^{TE} + {\mathbf{E}}^{TM}.
\end{equation}
The vector ${\vec{B}} $ and ${\vec{E}} $ must be solutions of
the Maxwell equations in the vacuum
\begin{eqnarray}\nonumber
\frac{1}{c} \frac{\partial {\bf B}}{\partial \,t} &=&-\nabla \wedge \,
 {\bf E}, \;  \; \;  \; \frac{1}{c} \, \frac{\partial {\bf E}}{\partial     \,t}=
\nabla \wedge \, {\bf B},
 \\
\label{Maxwell1}
\nabla \cdot {\bf B}&=&0, \; \; \; \; \nabla \cdot {\bf E}=0.
\end{eqnarray} 
The Maxwell equations (\ref{Maxwell1}) expressed in term of $B_r$ and of the coefficients $\mu$ and $\eta$ become
\begin{eqnarray}
\label{eq_Br}
(\frac{\partial^2}{\partial r^2} &+&\frac{2}{r^2} + \frac{4}{r} \frac{\partial}{\partial r}) B_r + \lapa B_r  -\frac{1}{c^2}  \frac{\partial}{\partial t}  B_r =0,
\\ \label{eq_mu}
\partial_{r^2}^2 \mu &+& \frac{2}{r} \partial_r \mu + \frac{1}{r^2} \lapa  \mu + k^2 \mu = \mu_S
\\ \label{eq_eta}
 \partial_{r^2}^2 \eta &+& \frac{2}{r} \partial_r \eta 
+ \frac{1}{r^2} \lapa \eta + k^2 \eta + \frac{2}{r^2} B_r  = \eta_S
\end{eqnarray}
where
\begin{eqnarray} \label{spherique_laplacien_angulaire}
\nabla^2_{\theta,\phi} &=& \partial_{\theta^2}^2 + \cot \theta \partial_{\theta} + \frac{1}{\sin^2 \theta} \partial_{\phi^2}^2 
\nonumber \\
&=& \frac{1}{\sin \theta} [\partial_\theta(\sin \theta \, \partial_\theta) +\frac{1}{\sin \theta} \partial_{\phi^2}^2]
\end{eqnarray}
is the angular Laplacian.
When it is time dependent, the electric field is deduced from $\eta$ and $\mu$ through the Faraday equation and the relations 
\begin{eqnarray}
(\nabla \times \vec B)_r &=& \frac{1}{r} \lapa \mu \nonumber \\
(\nabla \times \vec B)_\theta &=& \frac{1}{r \sin \theta}[\partial_\phi B_r - (I+r \partial_r) \partial_\phi \eta - \sin \theta (I+r \partial_r) \partial_\theta \mu] \nonumber \\
(\nabla \times \vec B)_\phi &=&  \frac{1}{r \sin \theta}[-\sin \theta \partial_\theta B_r 
+ \sin \theta (I+r\partial_r)\partial_\theta \eta
\nonumber \\ & &
 -(I+r\partial_r)\partial_\phi \mu].
 \label{pour_Faraday_mu_eta}
\end{eqnarray}
After separation of the variables, it is found that the angular solution of Eqs. \ref{eq_Br}-\ref{eq_mu} can be expanded in spherical harmonic functions $ Y^m_l(\theta, \phi) =P_l^m (\cos \theta) e^{i m \phi}$,
where $P^m_l(\cos \theta) $ are the associated Legendre functions.
The scalars  $ r$, $\theta $,
$\phi$, are the spherical coordinates the $z$ axis being the star spin axis. 

Two cases must be treated separately, depending on $m$. When $m=0$, the solution is axially symmetric, and not time dependent. The parts depending on $r$ in Eqs. (\ref{eq_Br}-\ref{eq_eta}) are simple differential equations with elementary solutions. When $m \ne 0$, the solution is time dependent, and the parts of Eqs. \ref{eq_Br}-\ref{eq_eta} that depend on $r$ can be converted into Bessel equations of the normalised variable $x= m\omega r/c$. 

We solve the $TM$ and $TE$ solutions separately.
The $TE$ solution is derived from $\eta$ and has a finite radial magnetic component $B_r$ given by Eq. (\ref{eq_Br}). This equation is solved directly (see the following sections). Then, using the divergence of the magnetic field  
\begin{equation} \label{spherique_potentiel_eta_divergence}
\nabla \cdot \vec B = \frac{\partial}{\partial r} B_r + \frac{2}{r} B_r +\frac{1}{r} \nabla^2_{\theta,\phi} \eta,
\end{equation}
 and the fact that  with a $Y_l^m$ angular dependence $\lapa \eta = -l(l+1) \eta$, we find $\eta$. Then, from  Eq. (\ref{def_eta_mu}), 
\begin{eqnarray} \nonumber
B_\theta &=& \frac{1}{l(l+1)} \left( x \frac{\partial^2 B_r}{\partial x \partial \theta} + 2 \frac{\partial B_r }{\partial \theta} \right), \\
B_\phi &=& \frac{ i m}{l(l+1) \sin \theta} \left( x \frac{\partial^2 B_r}{\partial x \partial \phi} + 2 \frac{\partial B_r }{\partial \phi} \right).
\end{eqnarray}

The $TM$ magnetic field is derived from $\mu$. The field $\mu$ is found directly by resolution of Eq. (\ref{eq_mu}).
 
The computation of the electric field is different in the cases $m=0$ and $m \ne 0$, which is detailed in sections \ref{sec_axial} and \ref{sec_non_axial}. 
The outgoing solution of the Maxwell Eqs.(\ref{Maxwell1})
must also satisfy the boundary conditions (BC)
\begin{equation}\label{B.C.} 
{\bf E}\times {\bf n} = \left[ \frac{1}{c} ({\bf \Omega} \times {\bf R})
\wedge {\bf B} \right] \times {\bf n}
\end{equation} 
at the surface of the star, where $ {\bf n} $ is the unit vector orthogonal
to the surface of the star, and $\bf{R} $ is the radial vector connecting the centre of the star to the point of interest on its surface.

Let be $R$ the radius of the spherical  neutron star (NS).
Inside of the NS $ r\le R$, the magnetic field is generated by 
internal currents. Let be ${\bf B}^{<}_{rlm}(r,\theta,\phi) $ the $l,m$ component
of the spectral decomposition of ${\bf B}$. The  electric field
${\bf E^{<}} $ inside the NS is
\begin{equation}
E^<_r = \Omega \frac{r}{c} B^<_\theta \sin \theta, \; \;
  E^<_\theta=-\Omega \frac{r}{c} B^<_r \sin \theta, \; \; E^<_\phi=0.
\end{equation} 
The matching conditions are (See Eq.(\ref{B.C.})) 
\begin{equation}\label{matchE}
B^>_r(R,\theta,\varphi)=B^<_r(R,\theta,\varphi), \; \;
E^>_\theta(R,\theta,\varphi)=E^<_\theta(R,\theta,\varphi),
\end{equation}\label{Brmatching}
and
\begin{equation}
E^>_\phi(R,\theta,\varphi)=0,
\end{equation} 
where ${\bf E}^> $ is the  field in the vacuum.

\section{Axially symmetric solutions and their matching conditions} \label{sec_axial}

In this section, we compute the multipole
electromagnetic field around a rotating neutron star satisfying axially symmetric
BC In terms of spherical harmonics, 
they correspond to $m=0$.
When $m=0$, there is a finite $TM$ solution derived from Eq. (\ref{eq_mu}), but the curl
of this magnetic field is finite too. This means that there is either a time varying electric field, or an electric current density. Because $m=0$, a time varying electric field is discarded. Since we are looking for a vacuum solution, a current density is discarded too. Therefore, only a $TE$ electromagnetic field is retained in the axially symmetric case $m=0$. 

Following the method exposed in section \ref{sec_methods}, it is found that
the components of the vacuum $TE$ magnetic field are:
\begin{eqnarray} \label{Bpar} 
B_r&=& B_{l}^0 \left(\frac{R}{r} \right) ^{l+2} \, P^0_l(\theta),
\nonumber
 \\ \nonumber
 B_{\theta}&=&-B_{l}^0
\frac{1}{l+1} \left( \frac{R}{r} \right)^{l+2} \frac{d \,
 P^0_l(\theta)}{d \, \theta}, 
\\ \label{B_symmetric}
 B_{\phi} &=&0,
\end{eqnarray} 
where $P^0_l(\theta)=\mathcal{P}^0_l(\cos \theta)$ and $\mathcal{P}^0_l$ is the Legendre polynomial of order $l$. 
{If the interior of the rotating star  is a perfect conductor, the internal electric field
 $\vec{E} $  vanishes in the co-rotating frame. Consequently, the
  electric field in the inertial frame is $\vec{E}^{<} =
  (\vec{\Omega}\land \vec{r} )\land \vec{B} $ and
\be\label{Epar}
E_r^<=+\Omega \frac{r}{c} \sin \theta \, B_{\theta}^<, \; \; \;
E_{\theta}^<=-\Omega
\frac{r}{c} \sin \theta B_r^<.
\ee
}
The value of $ B^{<}_{r}(R) $ , $E^{<}_{\theta} (R) $ and
$E^{<}_{\phi}(r) $ at the surface of the star $r=R$ determine the
boundary condition for the external field. 
Outside the star ( $ r \ge R$), the electric field must be the gradient of an
harmonic potential $\Phi$ {(No charge in the vacuum, steady magnetic field)} 
\begin{equation}
\Phi = \sum_{l'} \frac{C_{l'}}{r^{(l'+1)}} P_{l'}^0 (\theta)
\end{equation}
and its  component $E_\theta$ must match the
 components $E_{\theta} $ inside the star:
\be\label{match}  
E_{\theta}(R) = \frac{1}{R} \frac{\partial \, \Phi}{\partial \, \theta} =
-\Omega \frac{R}{ c } \sin \theta \, B_{l0}^0 \, P^0_l.
\ee 
This imposes a series of constraints on the coefficients $C_{l'}$.
The relation 
\be\label{dlegendre}
\frac{d}{d \theta}\left[ P^0_{l+1}(\theta) - P^0_{l-1}(\theta)
  \right]=-(2 l+1) P^0_l(\theta) \sin \theta
\ee   
is deduced from the derivative of Eq. 8.914.2 in \cite{Grad}   
and the differential equation defining the Legendre functions (Eq. 8.820, same reference). 
Equation (\ref{dlegendre}) is used to deduce the values of the $C_{l'}$ coefficients from Eq. (\ref{match}). Finally,
\begin{eqnarray} \label{Epar_BC}  \nonumber
\Phi &=& \frac{B_{l0}^0}{2 \, l+1} \left[  \left(\frac{R}{r} \right)^{l+2} P_{l+1}^0 (\theta) -\left(\frac{R}{r} \right)^{l} P_{l-1}^0 (\theta)\right] \Omega \frac{R^2}{c}  -\frac{Q}{r}\\  \nonumber
E_r&=& %-\frac{\partial \, \Phi}{\partial \,r} 
=\frac{B_{l0}^0}{2l+1} \left[
 ( l+2)  P^0_{l+1}\left( \frac{R}{r} \right)^{l+3} 
-l \, P^0_{l-1}  \left( \frac{R}{r} \right)^{l+1}
\right] \Omega \frac{R}{c} +\frac{Q}{r^2}
\\  \nonumber
E_\theta&=& % -\frac{1}{r}  \frac{ \partial \, \Phi}{\partial \, \theta} =
\frac{-B_{l0}^0}{2 \, l+1} \left[ \frac{d P^0_{l+1}}{d \theta}   
\left(\frac{R}{r} \right)^{l+3} - \frac{d P^0_{l-1}}{d \theta}
\, \left( \frac{R}{r} \right)^{l+1} 
\right] \Omega \frac{R}{c}
\\  
\label{E_axially_symmetric}
E_\phi&=&0.
\end{eqnarray}
 We have added in $\Phi$ and $E_r$ the effect of a possible global electric charge $Q$ of the NS.

\section{The non-axially symmetric solutions} \label{sec_non_axial}

The solutions corresponding to magnetic fields with an inclination $i=90^o$ over the $z$ axis correspond to $m \ne 0$.
They are developed in this section.

The $TE$ solution includes a magnetic field with a finite radial component $B_r$.
The solution of Eq. (\ref{eq_Br}) is 
\begin{equation} \label{spherique_alembert_forme_generale_solution}
B_r=  \sum_{l=1}^{\infty} \sum_{-l \le m \le l} \left[C_{lm}^{(1)}
  \frac{h_{l}^{(1)}(x)}{x} + C_{lm}^{(2)} 
\frac{h_{l}^{(2)}(x)}{x}\right] P_{lm}(\cos\theta) e^{i m (\phi-\Omega t)},
\end{equation} 
where $ \Omega= \| {\bf \Omega} \|$, $c$ is the light velocity,*
 $P_l^m(\theta)=\mathcal{P}_l^m(\cos \theta)$, and $\mathcal{P}_l^m$ is the associated Legendre polynomial of order $l,m$.
 The function 
$h_l(x)$ is the spherical Hankel function
\begin{equation}\label{Besel}
h_l(x) = \sqrt{\frac{\pi}{2 x}} H^{1}_{l+1/2}(x)
\end{equation}
where  $H^(1)_{l+1/2}(x) $ is the Bessel function of semi-integer order
$l+1/2$  and
\begin{equation}
x=m\, \frac{r \, \Omega}{c}, \; \; \; \varphi=\phi-\Omega \, t.
\end{equation}
$C_{lm}^{(1)}$ and $C_{lm}^{(2)}$ are constant numbers.
Because the solutions involving $h_l^{2}$  are associated with an incoming wave, we do not keep them, and for simplicity, we use the notation $h_{l}$ for  $h_{l}^{(1)}$.

For the $TM$ solution, $\mu$ is derived from Eq. (\ref{eq_mu}). Considering only a single $l, m$ term,
\begin{equation}
\mu= \sigma_{l,m} C_{lm}' h_l(x) Y_{lm} (\theta, \phi),
\end{equation}
where $C_{lm}'$ is a constant number.
The electric field is derived from the Faraday equation and Eqs. \ref{pour_Faraday_mu_eta}.
The solutions for the $TM$ and $TE$ components of the
electric and magnetic field are:
\begin{eqnarray}
&& B_{r;\, lm}^{TE} (r, \theta, \phi, t) =   A^{TE}_{lm}
 \ \frac{h_l(x)}{x}  \ P_l^{\, m  } (\theta) \
e^{i m \varphi}
\label{BrTElmP}
\\ \nonumber
&& B_{\theta;\, lm}^{TE}(r, \theta, \phi, t) =  \frac{A^{TE}_{lm}}{l(l+1)} 
\left(\frac{1}{x} \frac{d}{dx} \left(x \, h_l(x) \right) \right) 
\frac{d P_l^{\, m}  (\theta) }{d \theta} \ e^{im \varphi} 
\label{BthetaTEP}
\\ \nonumber
&&B_{\phi;\, lm}^{TE}(r, \theta, \phi, t) = i\, \frac{m A^{TE}_{lm}}{l(l+1)}  
\left(\frac{1}{x} \frac{d}{dx}\left(x \, h_l(x) \right) \right)
\frac{P_l^{\,  m }(\theta)}{\sin \theta} \ e^{im \varphi}
\label{BphiTEP}
\end{eqnarray}
\begin{eqnarray}
&&E_{r;\, lm}^{TE}(r, \theta, \phi, t) = 0
\\ \nonumber
&&E_{\theta;\, lm}^{TE}(r, \theta, \phi, t) =-  m \, A^{TE}_{lm} \, 
\frac{h_l(x)}{l(l+1)}\, \frac{P_l^{\, m} (\theta)}{\sin \theta} \ e^{im \varphi}
\label{EthetaTEP}
\\ \nonumber
&&E_{\phi;\, lm}^{TE}(r, \theta, \phi, t) = - i \, 
A^{TE}_{lm} \frac{h_l(x)}{l(l+1)} \,
\frac{d P_l^{\,m } (\theta)}{d\theta} \ e^{im \varphi}
\label{EphiTEP}
\end{eqnarray}
\begin{eqnarray}
&&B_{r;\, lm}^{TM}(r, \theta, \phi, t) = 0
\label{BrTMP}
\\ \nonumber
&&B_{\theta;\, lm}^{TM}(r, \theta, \phi, t) = +  m A^{TM}_{lm} \, 
\frac{h_l(x)}{l(l+1)} \, 
\frac{P_l^{\, m} (\theta)}{\sin \theta} \ e^{im \varphi}
\label{BthetaTMP}
\\ \nonumber
&&B_{\phi;\, lm}^{TM}(r, \theta, \phi, t) = +i \, 
A^{TM}_{lm} \ \frac{h_l(x)}{l(l+1)}  \
\frac{d P_l^{\, m}(\theta)}{d\theta} \ e^{im \varphi}
\label{BphiTMP}
\end{eqnarray}
\begin{eqnarray}
&& E_{r;\, lm}^{TM}(r, \theta, \phi, t) =
\,A_{lm}^{TM} \frac{h_l(x)}{x}\, P_l^{\, m}
 (\cos
 \theta)\, e^{im \varphi}
\label{ErTMP}
\\ \nonumber
&&E^{TM}_{\theta;\, lm}(r, \theta, \phi, t) = \frac{A^{TM}_{lm}}{l(l+1)} 
\left(\frac{1}{x} \frac{d}{dx}\left(x \, h_l(x) \right) \right)
\frac{d P_l^{\, m } (\theta)}{d\theta} \, e^{im \varphi} 
\label{EthetaTMP}
\\ \nonumber
&&E^{TM}_{\phi;\, lm}(r, \theta, \phi, t) = i\,  \frac{m\, A^{TM}_{lm}}{l(l+1)} 
\!\left(\frac{1}{x} \frac{d}{dx}\left(x \, h_l(x) \right) \right)
\!\!\frac{P_l^{\, m } (\theta)}{\sin \theta}\, e^{im \varphi}.
\label{EphiTMP}
\end{eqnarray}

\section{Matching conditions for the non-axially symmetric solutions} \label{sec_matching_conditions}

Let be $B^<_{r;\,lm}(R) P^m_l(\theta)e(i m \varphi) $ the $l,m$ component
of the internal field $B_r^<$ at the surface of the star.
Taking into account the elementary expression of the external
magnetic field given by Eq.(\ref{BrTElmP}) , the  
  matching conditions described by Eq.(\ref{Brmatching})
determine  the coefficient $A^{TE}_{lm} $ in 
Eq.(\ref{BrTElmP}). We have
\begin{equation}\label{ATEml}
A^{TE}_{lm}=B^<_{r;\,lm} (R)\frac{x_s}{h_l(x_s)}  
\end{equation}
where $x_s=m\Omega R /c$.
Note that the above B.C. is not sufficient to determine the
magnetic field uniquely: in fact, an arbitrary  toroidal component ${\bf B}^{TM}
$ defined by $B_r^{TM}=0 $ can be
added to the poloidal component
in a such a way that the electric counterpart ${\bf E}^{TM}$  allows us to satisfy the boundary conditions
 $$E^{TE}_{\phi}(R,\theta,\phi,t)+E^{TM}_\phi
(R,\theta,\phi.t) =0.  $$ This equation reads
\begin{eqnarray}\label{eqmaitresse}
&&
{\cal E}_{\phi}(R,\theta,\phi,t)=\sum_{lm}
- A^{TE}_{lm} \, \frac{h_l(x_s)}{l(l+1)} \,
\frac{d P_l^m(\theta)}{d\theta} \ e^{im\varphi}
 \\ \nonumber
&&+\sum_{l'm'} \
\frac{m'\, A^{TM}_{l'm'}}{l'(l'+1)} 
\left(\frac{1}{x} \frac{d}{dx}\left(x \, h_{l'}(x) \right) \right)_{r=R}
\!\frac{P_{l'}^{m'}(\theta)}{\sin \theta}\, e^{im'\varphi}  \ = \ 0.
\label{condEphinul}
\end{eqnarray}
The details of its resolution are given in Appendix \ref{sec_solution_Etheta_nul}. Only two coefficients 
$A^{TM}_{l'm'}$ remain in the right-hand side of Eq. (\ref{eqmaitresse}); they are 
\begin{equation}\label{TMMl1}
A^{TM}_{l+1 \, m}=A^{TE}_{l \, m} \frac{h_l(x_s)}{D_{l+1}} \frac{(l-m+1)(l+2)}{m(2l+1)} 
\end{equation}
and
\begin{equation}\label{ATMlmoins1}
A^{TM}_{l-1 \, m}= - A^{TE}_{lm} \,  \frac{h_l(x_s)}{D_{l-1} } \frac{(l+m)(l-1)}{m(2l+1)},
\end{equation}
where the coefficient $D_l$ is defined in Eq. (\ref{eq_D_l}).
Finally, the $r$ and $\theta$ component of the total
electric field $ {\bf {\cal E}}_{lm} $ are:
\begin{eqnarray} \nonumber
{\cal E}_{r,lm}(r,\theta,\phi)&=& A^{TM}_{l+1 \,m} \frac{h_{l+1}(x)}{x}
P^m_{l+1}(\theta)e ^{i \, m \varphi}
\nonumber \\ \label{bolEr} \nonumber
&& +A^{TM}_{l-1 \,m} \frac{h_{l-1}(x)}{x} \,P^m_{l-1}(\theta)
 \, e ^{i \, m \varphi},
\end{eqnarray}   
\begin{eqnarray}\label{bolEthet} \nonumber
 {\cal E}_{\theta,lm}&=&-m A^{TE}_{lm} \frac{h_l(x)}{l(l+1)}
  \frac{P^m_l(\theta)}{\sin \theta} \, e^{i m \varphi}  
\\
\nonumber
&& +\frac{A^{TM}_{l+1 \, m}}{(l+1)(l+2)} \left( \frac{1}{x} \frac{ d}{dx}(x
  h_{l+1}(x)) \right) \frac{d P^m_{l+1}(\theta)}{d \theta} \, e^{i\, m \varphi} 
\\
\nonumber
&& +\frac{A^{TM}_{l-1 \, m}}{l (l-1)} \left( \frac{1}{x} \frac{d}{dx} (x h_{l-1}(x)) \right)
\frac{d P^m_{l-1}(\theta)}{d \theta} \, e^{i m \, \varphi},
\end{eqnarray}
\begin{eqnarray}\label{E_asymmetric}
 {\cal E}_{\phi, lm}&=&-i \, A^{TE}_{lm} \frac{h_l(x)}{l (l+1) }
  \frac{d P^m_l(\theta)}{ d \theta}  \, e^{i \, m \varphi}
\\
\nonumber
&& + i \, \frac{m A^{TM}_{l+1 \, m}}{(l+1)(l+2) } \left( \frac{1}{x}
  \frac{d}{dx}(x h_{l+1}(x)) \right) \frac{P^m_{l+1}(\theta)}{\sin \theta}
\, e^{i m \varphi}
\\
\nonumber
&& +i \frac{m \, A^{TM}_{l-1 m}}{l (l-1)} \left( \frac{1}{x}
  \frac{d}{dx} (x h_{l-1}(x)) \right) \frac{P^m_{l-1}(\theta)}{\sin
  \theta} \, e^{i \, m \varphi}.
\end{eqnarray}
The electric field computed above, satisfies the
boundary condition ${\cal E}_{\phi,lm}(R,\theta, \phi ,t) $ =0. 
It is shown in section \ref{matching_condition_is_ok} that they also fit the boundary condition given by Eq.(\ref{matchE}) 

The magnetic counterpart ${\cal B}$  is given by (See
Eq. \ref{BrTMP}, Eq. \ref{BthetaTMP}, Eq. \ref{BphiTMP})
\begin{equation*} \label{bolBr} 
 {\cal B}_{r,lm}(r,\theta,\phi,t)=A^{TE}_{lm} \frac{h_l(x)}{x}
\, P^m_l(\theta) \, e^{i \, m \varphi},
\end{equation*}
\begin{eqnarray}\label{bolBtheta} \nonumber
{\cal B}_{\theta,lm}(r,\theta,\phi,t)&=&\frac{A^{TE}_{lm} }{l \,(l+1)}
\left( \frac{1}{x} \frac{d}{dx} (x h_l))  \right)  \frac{d  P^m_l ( \theta)}{d \theta}
\,e^{i \, m \varphi}
\\
\nonumber
&& +A^{TM}_{l+1 \, m} \frac{m}{(l+1)(l+2)} h_{l+1}(x) \frac{P^m_{l+1}(\theta)}{\sin\theta} 
\, e^{i \, m \varphi}
\\
\nonumber
&& + A^{TM}_{l-1 \, m} \frac{m}{l \, (l-1)} h_{l-1}(x) \frac{P^m_{l-1}(\theta)
}{\sin \theta} \, e^{i \, m \varphi},
\end{eqnarray}
\begin{eqnarray} \label{B_asymmetric}
 {\cal B}_{\phi,lm}(r,\theta,\phi,t) &=& i \frac{m A^{TE}_{lm}}{l \, (l+1)}  
\left( \frac{1}{x} \frac{d}{dx} (x h_l(x))\right)
\frac{P^m_l(\cos \theta)}{\sin \theta} \,  e^{i m \varphi}
\\
\nonumber
&& +i  \, A^{TM}_{l+1,m} \frac{h_{l+1}}{(l+1)(l+2)} \, \frac{ d \,
  P^m_{l+1}(\cos \theta)} {d \, \theta} \, e^{i \, m \varphi} 
\\
\nonumber
&& +i  A^{TM}_{l-1,m} \frac{h_{l-1}(x) }{(l-1)\,  l} 
\, \frac{d \, P^m_{l-1}(\cos \theta)}{d \theta} \, e^{i \, m \varphi}.
\end{eqnarray} 

For $l=1$, $m=1$ we obtain the result given in \cite{Deutsch_1955}.

\section{A pulsar that extracts electrons from one pole and protons from the other} \label{sec_extracts_two_kinds_of_charges}

 With dipole pulsar magnetosphere, the open field lines above the two opposite poles present vertical electric fields and Goldreich-Julian currents
 of the same sign. Therefore, the particles that are extracted from the two poles of the neutron star have the same electric charge. 
  With  pulsar dipole magnetosphere model ending with a wind, there is a continuous flux of emitted particles, and it is necessary to close the currents, otherwise the neutron star would accumulate electric charges.  Charge accumulation cannot be indefinite, and it is generally assumed that the wind particles (of both positive and negative charges) come from pair creations. The pairs need a continuous flux of primary particles, however,  and the question of charge neutrality, i.e. current closure, remains with the primary particles. 
  
Static pulsar electrospheres \citep{Petri_2002} are models that do not involve charge circulation. Unfortunately, they do not create a wind either, and they are not expected to radiate. Aligned electrospheres have a dome of charged particles of one sign above each pole, and an equatorial belt of particles of the opposite sign. In that configuration, a  dicotron instability can develop. The dicotron effect tends to modulate the shape of the equatorial belt, and it can expel some of its particles \citep{Petri_2002b}. Then, particles of the two signs can be ejected from the neutron star, and this solves the problem of charge neutrality and current closure. 
 
In the present section, we present an alternative to electrospheres and dicotron instability that solves the charge neutrality problem. It consists of a neutron star with a multipole magnetic field. For simplicity, we consider only an aligned dipole and a quadrupole component.

\begin{figure}
%\resizebox{\hsize}{!}{\includegraphics{fig_ailealffig1bis.eps}}
\resizebox{\hsize}{!}{\includegraphics{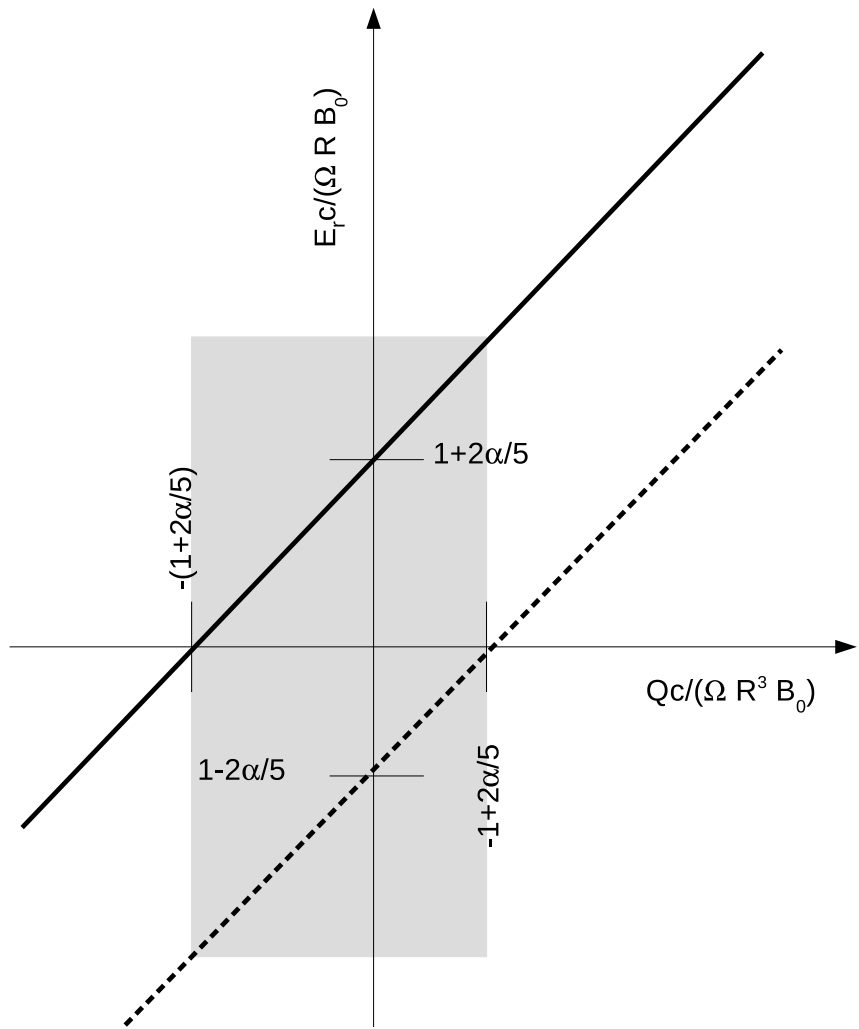}}%{fig_extracts_two_kinds_of_charges.eps}}
\caption{Normalised radial electric field $E_r$ as a function of the normalised electric charge $Q$ at the north pole (thick continuous line) and south pole (thick dashed line). The grey area represent the domain where particles of opposite charges can be extracted from opposite poles. }\label{fig_extracts_two_kinds_of_charges}
\end{figure}

\begin{figure}
%\resizebox{\hsize}{!}{\includegraphics{fig_ailealffig1bis.eps}}
\resizebox{\hsize}{!}{\includegraphics{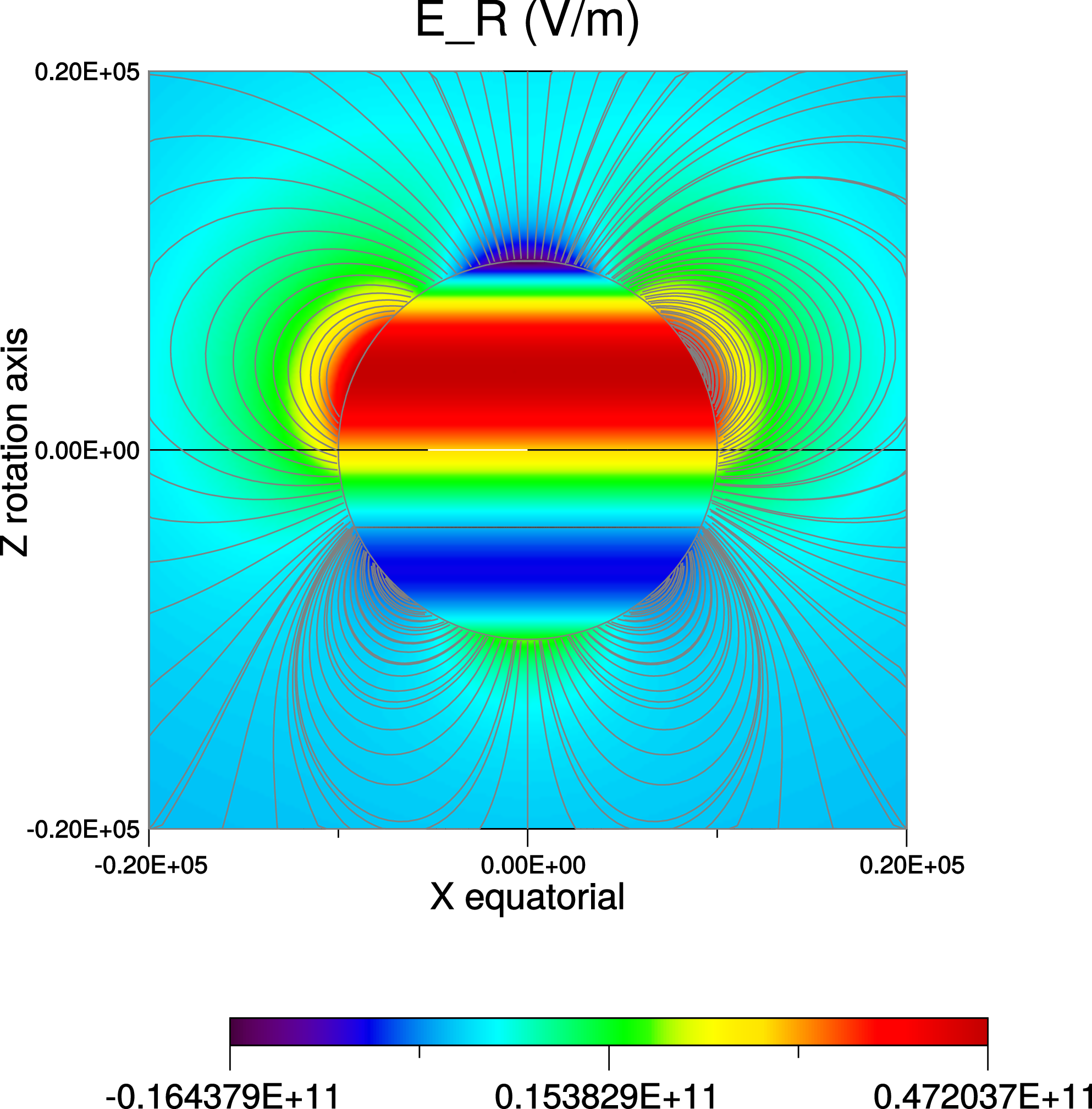}}%{fig_axial_Er_charge_positive.ps}}
\caption{A dipole field  $A^{TE}_{  1,  0}=$-1, and a quadrupole field $A^{TE}_{  2,  0}=$-2.5, and a total electric charge $Q=0.5$. The colour code represents the radial electric field $E_r$ plotted on the NS surface (within the circle that delimits the surface) and in a meridian plane perpendicular to the line of sight  (outside the circle that delimits the NS surface). Magnetic field lines with a foot on the surface in the same meridian plane are plotted as well.}\label{fig_axial_Er_charge_positive}
\end{figure}

\begin{figure}
%\resizebox{\hsize}{!}{\includegraphics{fig_ailealffig1bis.eps}}
\resizebox{\hsize}{!}{\includegraphics{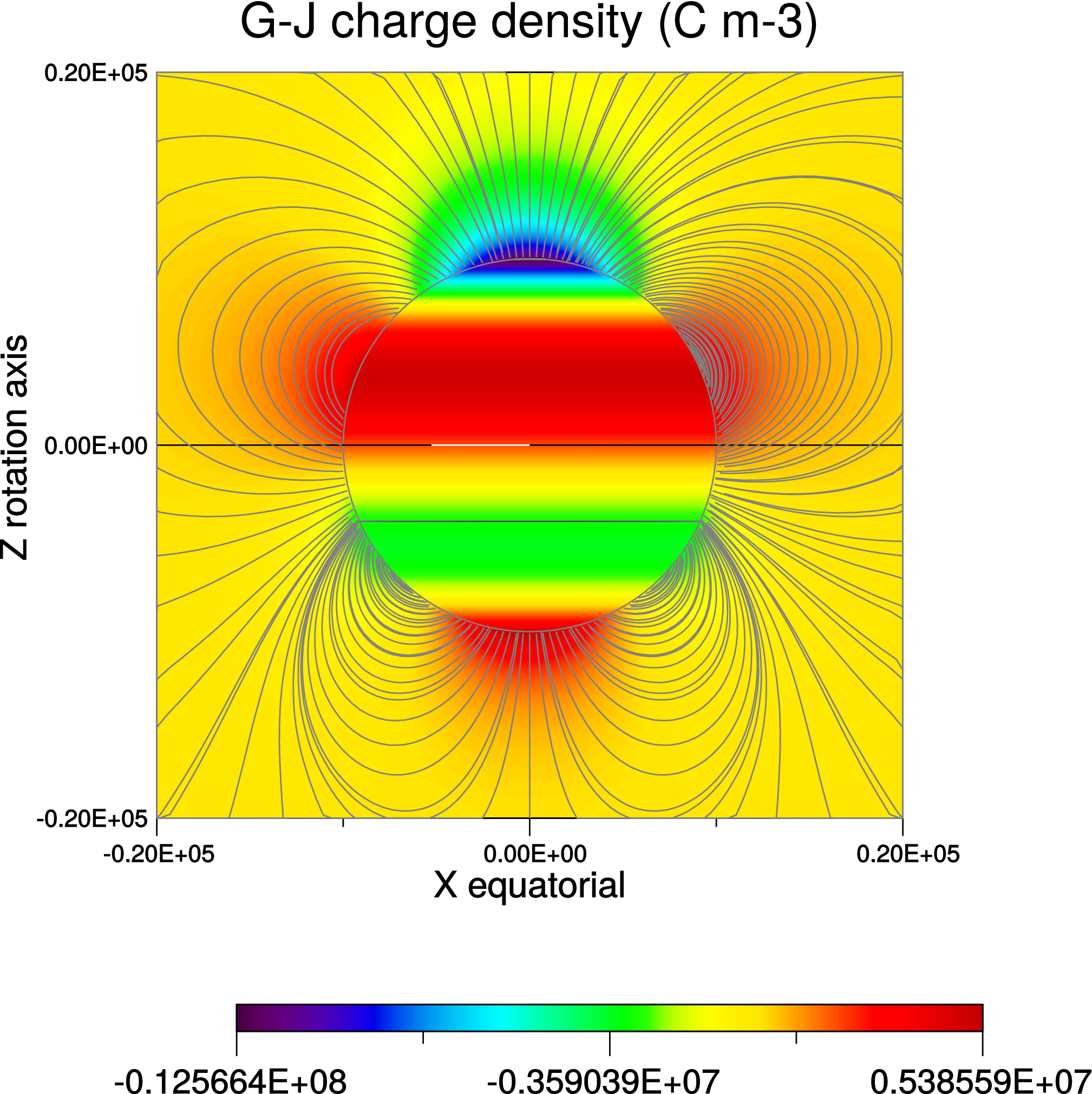}}%{fig_axial_NGJ_charge_positive.ps}}
\caption{Goldreich-Julian density $n_{GJ}$ with the same mutipole components as in Fig. \ref{fig_axial_Er_charge_positive}.}\label{fig_axial_NGJ_charge_positive}
\end{figure}

We have
\be\label{Brmaosph}
B_r=B_0 \left[ \left( \frac{ R}{r}\right)^3 \cos \theta + \frac{\alpha}{2}
  (3 \cos \theta^2 -1) \left( \frac{R}{r} \right)^4
\right]
\ee
\be\label{Bthetmagn}
B_\theta=B_0\left[\frac{1}{2} \left(\frac{R}{r}^3 \right)^3 \sin \theta 
+\alpha \left( \frac{R}{r} \right)^4 \cos\theta \sin \theta 
\right]
\ee
where $\alpha$ characterises the quadrupole component amplitude.

The radial electric field is
\begin{eqnarray}
\label{Ermag}
E_r &=& +\tilde{\Omega} B_0 \left[\frac{1}{2}( 3 \cos \theta^2-1)
 \left(\frac{R}{r} \right)^4 \right]+\frac{Q}{R^2}\left(\frac{R}{r} \right)^2
\\ \nonumber
&+&\tilde{\Omega} B_0 \left[+\frac{2}{5}\alpha \left(  (5 \cos \theta^3 -3 \cos \theta \, ) \left(
\frac{R}{r} \right)^5
- \cos \theta \left( \frac{R}{r} \right)^3 \right)  \right]
\end{eqnarray}
where $\tilde{\Omega} = \Omega \, R/c $ and $Q$ is an integration
constant depending on the total charge of the star.

In what follows, we consider a dynamical process: the electrons are extracted and accelerated from the north pole  
($\tilde{\Omega} B_0<0$) and  at time $t= 0$, $Q(t)=0$.

Electrons are accelerated above the star surface
and they create  electron-positron pairs. The vacuum
electric field is then progressively screened by the pairs. 
If the protons remain attached in the vicinity of the star, the magnetosphere charges as long as the electrons
are extracted and accelerated. Then, the total electric charge $Q$ of the star increases. Figure \ref{fig_extracts_two_kinds_of_charges} illustrates the evolution
of the radial electric field at the two poles. It is shown that if $Q$ increases (beware of signs, the normalised charge decreases) the radial electric fields on the north pole is less negative, and those on the south pole becomes more positive. Provided that $\alpha > 5/2$,  a finite range of values of $Q$ (highlighted by a grey rectangle) allows for radial electric fields of opposite signs at the opposite poles. 

Fig. \ref{fig_axial_Er_charge_positive} shows a  
 numerical example of superimposed and aligned dipole and quadrupole fields. 
   The only finite multipole components are characterised by the coefficients (here purely real)
 $A^{TE}_{  1,  0}=$-1, and $A^{TE}_{  2,  0}=$-2.5 and the total electric charge is $Q=0.5$ C.
  The electromagnetic field is computed on a spherical grid 
 extending from the star surface to a distance of 716.2 star radii 
 (15 light cylinder radii). The figure only represents  the area very close to the star, where the 
 quadrupole component is noticeable.
 The magnetic field has the intensity   $10^5$ T on the surface,
 and the dipole angle with the rotation axis is  null.
 The period of rotation of the star is      10 ms,
 which corresponds to a rotation frequency    628 s$^{-1}$
  and to a light cylinder radius  $0.47 \, 10^6$ m.
  We can see (colour code) the radial electric field $E_r$ on the left-hand side
as well as magnetic field lines.
  Because of the quadrupole component, the radial electric field does not have the same value on the two poles. 
  Its high negative value on the north pole is appropriate for the acceleration of electrons out of the star. On the north pole, the positive electric field can accelerate positive ions.

In comparison to $Q=0$ (not shown on a figure), the radial electric field amplitude with $Q=0.5$ is  
reduced (but still negative) in the north pole and more positive in the south pole, where protons can be accelerated. Then, the ability of the proton
to create pairs is increased, while those of the electron to create pairs remain high.
When pairs are created above the two opposite poles, a stationary regime is attained where both electrons and protons are extracted from the star, 
the total charge reaches an asymptotic value $Q_{0}$, and the pulsar can be active. 
Of course, in this regime, and especially if the NS surface is hot, the Goldreich-Julian density is an important marker of primary charge extraction.
We can see on Fig. \ref{fig_axial_NGJ_charge_positive} that the Goldreich-Julian density $n_{GJ}=\nabla \cdot \left( \vec{B}
\times \vec{V}_\Omega \right)/4 \pi \,c$ also has opposite signs at  opposite poles ($V_{\Omega}$ is the rotational velocity).

Of course, above the pair creation fronts, the electromagnetic field cannot correspond
to the vacuum model derived in this paper. But this model is useful below the pair creation front, where the flux of primary particles is not expected to induce currents that could significantly change the magnetic field topology. 

With this example, we do not argue that multipole fields are the most common solution to the pulsar current closure problem, but they represent at least one possibility.

At the opposite limit to vacuum approximation, the force-free equations of a magnetosphere were solved in a way that resolves the current system closure. This was done in 2D for an axially symmetric pulsar magnetosphere 
\citep{Contopoulos_1999,Gruzinov_2005, Gruzinov_2007}
and  for a 3D  dissipative force-free magnetosphere where the magnetic axis is not necessarily aligned with the rotation axis
\citep{Spitkovsky_2006, Kalapotharakos_2009}. Those models are based on a dipole magnetic field at the NS surface. The current closes through an equatorial current sheet where the current is opposite to that carried in the open field lines regions. Since force-free equations do not include the plasma transport equations (no explicit equation of density and momentum, for instance), the force-free models do not say much about the nature of the particles  that carry currents. It is generally argued that the equatorial return current is carried by electrons that were launched in open field line regions, as well as by positrons moving to the opposite direction, which result from pair creation cascades initiated by primary accelerated electrons. 

At a large distance from the NS, a vacuum solution associated with a multipolar electromagnetic field is not different from that associated with a dipole 
field. This probably holds with a plasma filled magnetosphere. Force-free magnetosphere associated with a surface multipole field might be very analogous to those with dipole fields at distances larger than a fraction of the light-cylinder radius, but the current sheet could be different near the star. As we will see in the next section, this can affect the pulse shape.

\section{Pulse shape} \label{pulse_shape}

\begin{figure}
%\resizebox{\hsize}{!}{\includegraphics{fig_ailealffig1bis.eps}}
\resizebox{\hsize}{!}{\includegraphics{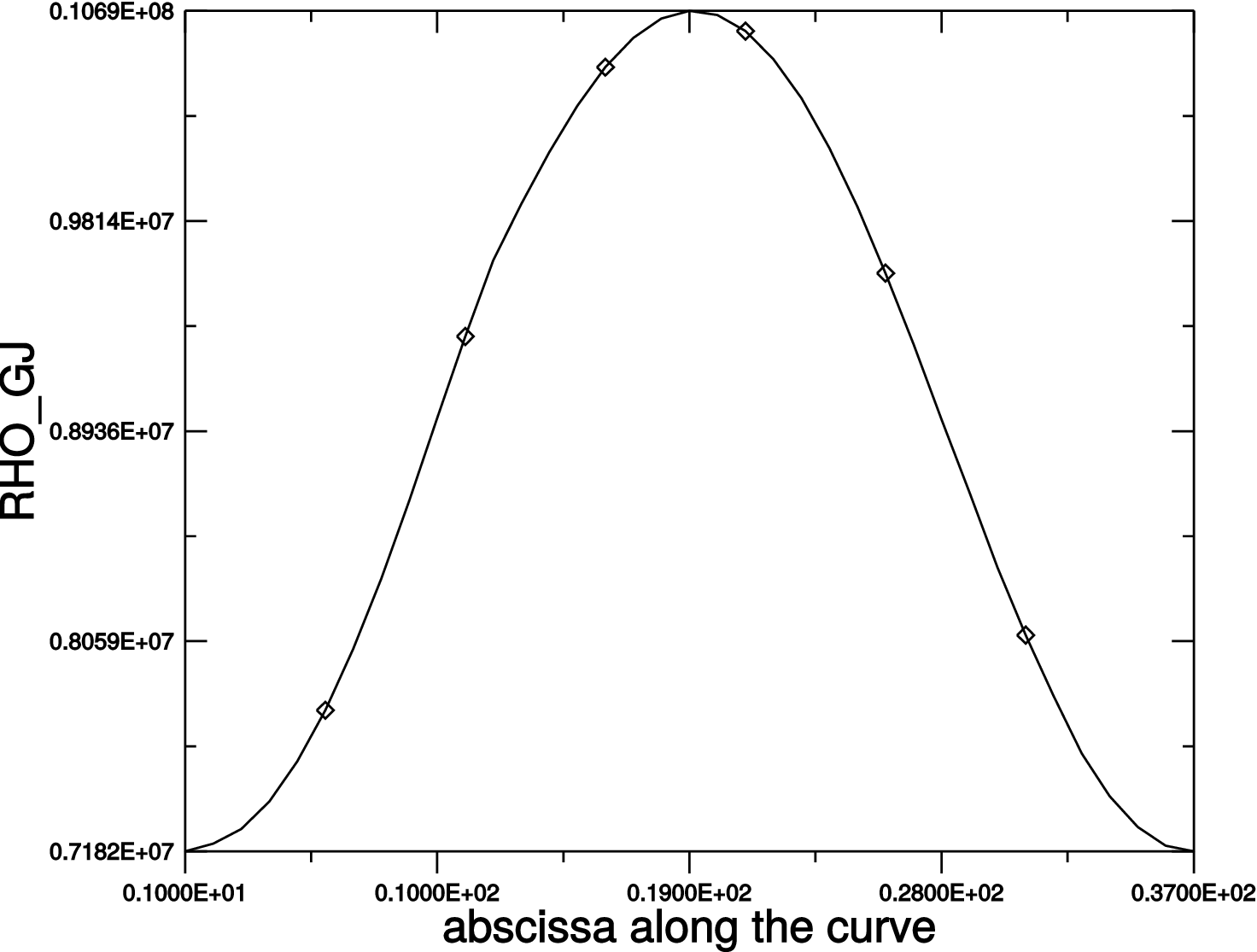}}%{fig_NGJ_dipole_profil.ps}}
\caption{Values of the Goldreich-Julian density on the NS surface at the foot of the last open magnetic field lines for a dipole magnetic field of inclination $i=40$ deg.}\label{fig_NGJ_dipole_profil}
\end{figure}

\begin{figure}
%\resizebox{\hsize}{!}{\includegraphics{fig_ailealffig1bis.eps}}
\resizebox{\hsize}{!}{\includegraphics{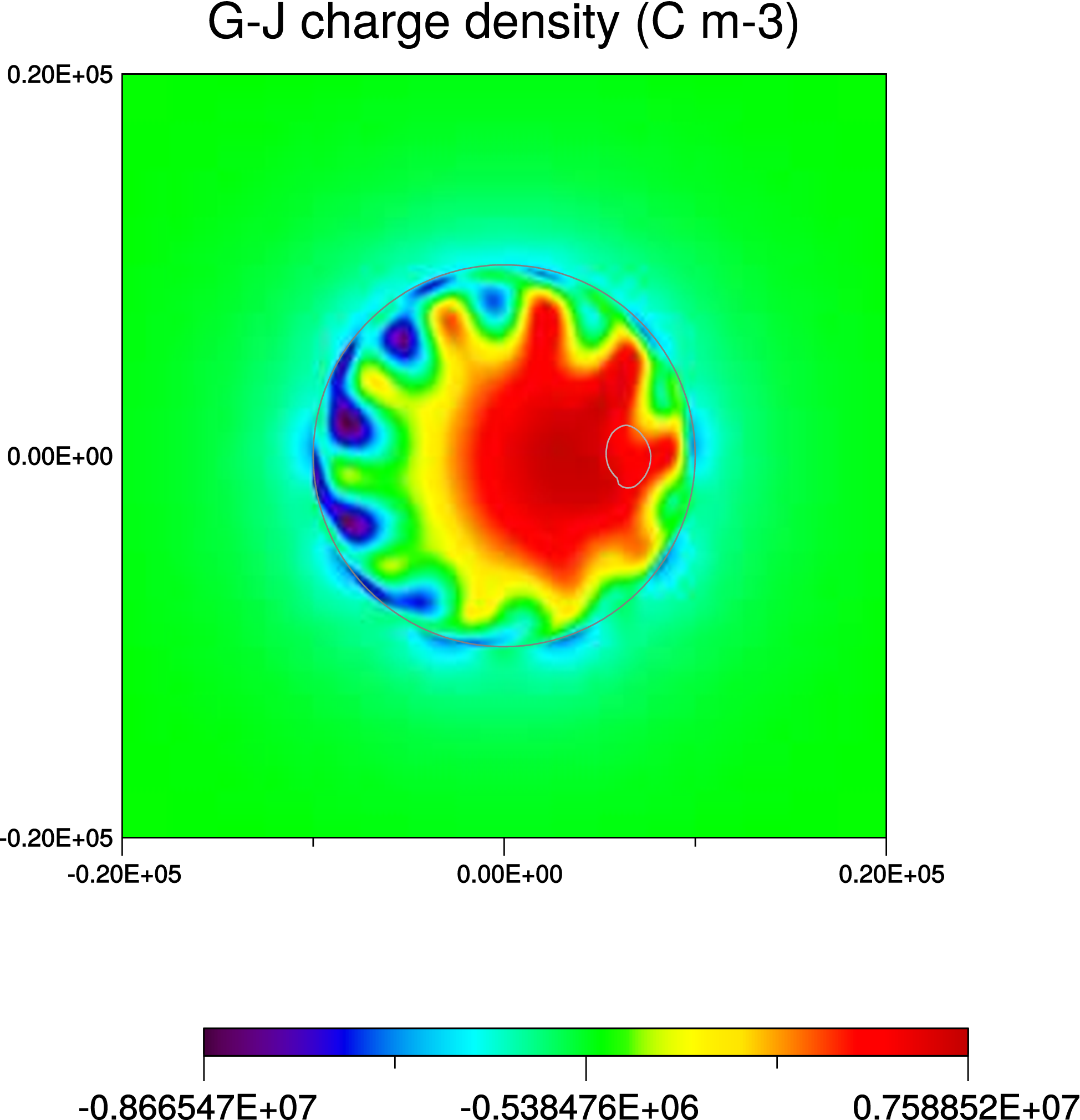}}%{fig_NGJ_multipole.ps}}
\caption{Values of the Goldreich-Julian density on the NS surface at the foot of the last open magnetic field lines for the multipole magnetic field described in Table \ref{multipole_exemple_pulse_shape}. The line drawn on the star surface corresponds to the foot (on northern hemisphere) of the last open field lines. }\label{fig_NGJ_multipole}
\end{figure}

\begin{figure}
%\resizebox{\hsize}{!}{\includegraphics{fig_ailealffig1bis.eps}}
\resizebox{\hsize}{!}{\includegraphics{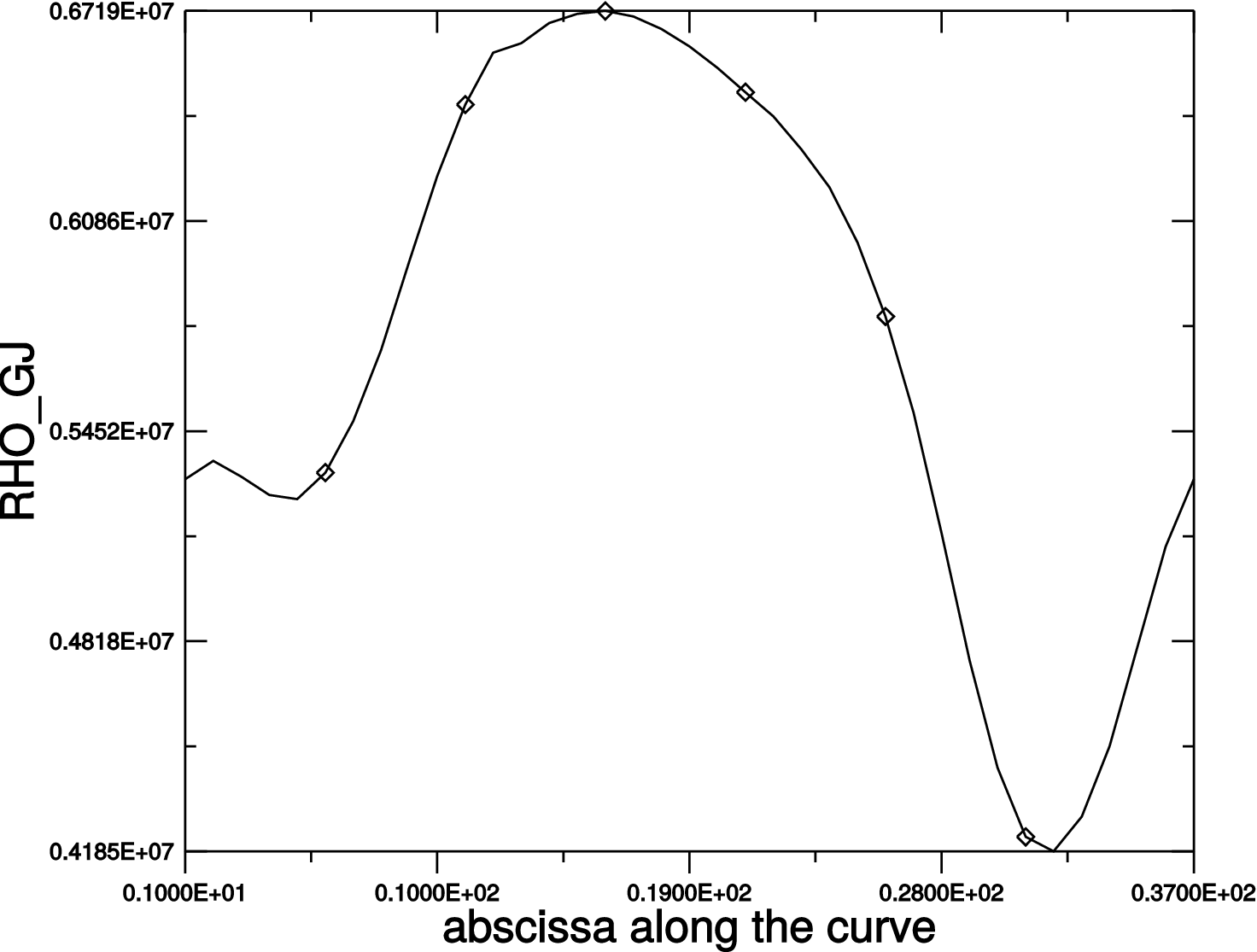}}%{fig_NGJ_multipole_profil.ps}}
\caption{Values of the Goldreich-Julian density on the NS surface at the foot (on northern hemisphere) of the last open magnetic field lines for the same multipole as in Fig. \ref{fig_NGJ_multipole_profil}.}\label{fig_NGJ_multipole_profil}
\end{figure}

In the standard model of the magnetosphere, the strong electric field
at the surface of the star $r=R$ extract and accelerate electrons from
the crust at relativistic energy. The current density  is
$J \sim e n_{GJ} c$ where $n_{GJ}$ is the Goldreich-Julian density.
Those primary electrons follow the lines of the magnetic field
$\vec{B} $  radiate high-energy $\gamma$ rays via curvature
radiation, and the gamma rays produce electron positron pair via the
magnetic field $\vec{B} $, if $\vec{B} $ is strong enough, or by $
\gamma$  rays and  crust thermal background $x$-ray mechanism. 
Electron positron pairs are supposed to generate the observed radio high
energy emission. The main consequence of this mechanism is that
the observed pulse shapes depends strongly on the Goldreich-Julian
density at the surface of the star. 

As mentioned in the introduction, the most often invoked heuristic models to explain pulse shapes are the polar cap model, 
the slot gap and caustics models, and the outer gap model.
In most of these models,  a critical area where $n_{GJ}$ determines the pulse shape is the curve drawn on the NS surface 
that corresponds to the feet of the last open field lines. 
Figure \ref{fig_NGJ_dipole_profil} shows $n_{GJ}$ at the NS surface at the feet of the last open magnetic field lines for a dipole field with an inclination $i=40$ deg. 
Its variations are very simple, symmetric, with a single maximum and a single minimum. 
Multipole components are now added to this dipole field. Their coefficients are displayed in Table \ref{multipole_exemple_pulse_shape}. The Goldreich-Julian density $n_{GJ}$ on the NS surface 
is displayed in Fig. \ref{fig_NGJ_multipole}. We can see the inclined dipole structure and the superimposition of smaller scale structures with a significant azimuthal modulation.
The line corresponding to the feet of the last open field lines is displayed (for the northern hemisphere). The values of $n_{GJ}$ are displayed in Fig. \ref{fig_NGJ_multipole_profil} as a function of the abscissa along the line. We can see that it is more complex than the dipole profile in Fig. \ref{fig_NGJ_dipole_profil}. The curve has secondary extrema and it shows a higher range of values. Without entering into the detail of pulse shape theories (it is not the topics of the present paper),  we can expect that the multipole field  can be associated with irregular pulse shapes  like those displayed in Figs. \ref{fig_psr_binaire_A_radio_Kramer_2006} and \ref{fig_pulse_vela_fermi_2009}. 

\begin{table}
\caption{Complex values of the finite $A^{TE}_{l,m}$ coefficients for the multipole field used as an example in section \ref{pulse_shape}. For other values of $l,m$, the coefficients are null.}
\label{multipole_exemple_pulse_shape}
\begin{center}
\begin{tabular}{|l|l| c|} % centered columns (4 columns)
 \hline 
 $l$  & $m$  & $A^{TE}_{l,m}$ \\
 \hline 
   1  &  0   & 0.766E+00$+i$  0.000E+00,\\
   1  &  1   & 0.643E+00$+i$  0.000E+00,\\
  11  &  9   & 0.255E-24$+i$ -0.126E-23,\\
  11  & 10   & 0.456E-24$+i$ -0.120E-23,\\
  12  &  7   & 0.594E-27$+i$ -0.246E-27,\\
  12  &  8   & 0.128E-26$+i$ -0.630E-26,\\
  12  &  9   & 0.128E-26$+i$ -0.630E-26,\\
  12  & 10   & 0.600E-26$+i$ -0.230E-26,\\
 \hline %inserts single line
\end{tabular}
\end{center}
\end{table}

\section{Conclusion}

We have developed an analytical formalism allowing us
the most general solution for an electromagnetic field in vacuum 
fulfilling the boundary conditions on the surface of a rotating magnetised star. 
This solution, based on an expansion on spherical harmonics is the linear combination of two types of contributions : 
 axially symmetric fields (azimuthal number $m=0$) given by Eqs. 
Eq. (\ref{B_symmetric}, \ref{E_axially_symmetric}), and non-axially symmetric fields ($m \ne 0$) given by Eq. (\ref{E_asymmetric}, \ref{B_asymmetric}).

 Of course, NS are well known to extract plasma in their immediate vicinity, therefore this solution cannot be used as is. 
 Nevertheless, 
 we showed in section \ref{sec_extracts_two_kinds_of_charges} that the presence of a
 quadrupole component of the magnetic field can solve the problem of
 the current closure in the pulsar magnetosphere. 
 As suggested in section \ref{pulse_shape}, this formalism can also be useful in modelling the observed pulse shapes in
 pulsars emission.
 
 This solution can be used as a benchmark for codes solving the electromagnetic field equations in the surrounding of a rotating magnetised star.

\cite{Petri_2013} has built numerical solutions of the electromagnetic field surrounding a star with a dipole field in the context of general relativity. 
Our model does not include gravitational effects, but it is possible that a numerical solution can be developed as well. The present solution can be used as a 
test when strong gravitational effects are neglected. 
 
  Moreover, the vacuum
electromagnetic solution can be the first step in an iterative
process to find more suitable pulsar models, where a plasma is (numerically) progressively introduced into the NS environment.

\appendix
\section{Derivation of the $A_{l'm'}^{TM}$ coefficients} \label{sec_solution_Etheta_nul}
The coefficients $A^{TM}_{l^{`} m^{'}} $ are computed by taking  the proprieties of the associated Legendre functions into
account. We obtain:
(Eq. 8.733,1 in Gradshteyn et al. 2007% \citep{Grad}
) 
\begin{equation}
- \sin^2 \theta \, 
\frac{d P_l^m(\theta)}{d (\cos \theta)} = l \, \cos \theta \ P_l^m(\theta) - (l+m) \ P_{l-1}^m(\theta)
\end{equation}  
or equivalently
\begin{equation}\label{DPlm}
+ \sin \theta \ \frac{d P_l^m(\theta)}{d \theta } = l \,  \cos \theta \ P_l^m(\theta) \, - (l+m) \ P_{l-1}^m(\theta).
\end{equation}
By using the expression (Eq.(8.731,2) in \cite{Grad})
\begin{equation}
(2 \,l+1) \, \cos \theta \, P^m_l=(l-m+1) P^m_{l+1} + (m+l) P^m_{l-1},
\end{equation}
and Eq.(\ref{DPlm}) reads
\begin{equation}\label{reladPPsursin}
\sin \theta \ \frac{d P_l^m}{d \theta } = \frac{l\, (l-m+1)}{2l+1} \ P_{l+1}^m 
- \frac{(l+1)(l+m)}{2l+1} \ P_{l-1}^m.
\end{equation}
By replacing the above value of $\sin \theta \, d\,P^m_l(\cos \theta) /d \, \theta $ in 
Eq.(\ref{eqmaitresse}) 
we obtain
\begin{eqnarray}
&& A^{TE}_{lm} \frac{h_l(x_s)}{l \,(l+1)}\, \frac{1}{(2l+1)} \left[\, l \,(l-m+1)
  P^m_{l+1}-(l+1)(l+m) P^m_{l-1} \right]  
\nonumber  \\ \label{eqmaitresse_2}
&& = \sum_{l^{'},m^{'}}  \frac{
m^{'} A^{TM}_{l{'} m^{'}}}{l^{'}(l^{'}+1)} D_{l^{'}}
  P^{m^{'}}_{l^{'}} e ^{i \, (m^{'}-m) \phi},   
\end{eqnarray}
where 
\begin{equation} \label{eq_D_l}
D_{l^{'}}=\left[ \frac{1}{x} \frac{d}{dx}(x h_{l^{'}} (x))\right]_{r=R}.
\end{equation}
  By multiplying both sides of Eq.({\ref{eqmaitresse_2}) by $\sin
  \theta \,  P^m_{l+1} (\cos \theta) \, e^{-i m\varphi}$,
after the  integration on $\theta$ between  $ 0$ and $\pi$, and on  $\phi$
between $0$ and  $2\pi$, with 
  Eq.(\ref{reladPPsursin}) and  the orthogonality properties of the
  associated Legendre functions  $P_l^m(\theta)$,
\begin{equation}
\int_0^\pi P_l^m\, P_k^m \sin \theta d
\theta=\delta_{l,k}\,\frac{2}{2l+1} \frac{(l+m)\, ! }{(l-m)\, !}
\end{equation}
%###
only two coefficients survive,we obtain Eq. (\ref{TMMl1}). 
Equation \ref{ATMlmoins1} is derived in an analogous way.

\section{Proof that the last boundary condition is fulfilled} \label{matching_condition_is_ok}
We obtained a solution that fulfils the condition $E_\phi(R)=0$. Does it fit the last 
condition imposed by the boundary condition $E_\theta(R)=-(r \Omega /c) B_r \sin \theta$ ?
From Eqs. \ref{BrTElmP}-\ref{EphiTMP}, this requirement is equivalent to 
\begin{eqnarray}
&&-m A_{lm}^{TE} \frac{h_l(x_s)}{l(l+1)} \frac{P_l^m (\cos \theta)}{\sin \theta}
\nonumber \\ 
&&+
A_{l+1 \, m}^{TM} \frac{D_{l+1}}{(l+1)(l+2)} \frac{d P_{l+1}^m (\cos \theta)}{d \theta}
+
A_{l-1 \, m}^{TM} \frac{D_{l-1}}{l(l+1)} \frac{d P_{l-1}^m (\cos \theta)}{d \theta}
\nonumber \\ 
&& \hspace{1cm}=
-\frac{1}{m} A_{lm}^{TE} h_l(x_s) P_l^m (\cos \theta) \sin \theta.
\end{eqnarray}
When the coefficients $A_{l+1 \, m}^{TM}$ and $A_{l-1 \, m}^{TM}$ are expressed using Eqs. \ref{TMMl1} and \ref{ATMlmoins1}, the requirement becomes
\begin{eqnarray} \label{condition_bord_1}
0=\left[-\frac{m}{l(l+1) \sin \theta} + \frac{\sin \theta}{m} \right] P_l^m(\theta)
+\frac{1}{m l (l+1)} \times
 \nonumber \\ \nonumber
\left[ \frac{l(l-m+1)}{(2l+1)} \frac{d P_{l+1}^m(\theta)}{d \theta}
-\frac{(l+m)(l+1)}{(2l+1)}\frac{d P_{l-1}^m(\theta)}{d \theta}\right].
\end{eqnarray}
Considering the derivative of Eq. (\ref{reladPPsursin}) relatively to $\theta$, 
the condition becomes
\begin{equation} \label{condition_bord_1_bis}
\left[-\frac{m}{l(l+1) \sin \theta} + \frac{\sin \theta}{m} \right] P_l^m(\theta)
+
\frac{d}{d \theta}\left[ \sin \theta \frac{d P_l^m}{d \theta}\right].
\end{equation}
By definition, the Lagrange polynomials $P_l^m$ are the solutions of the differential 
equation 
\begin{equation}
\frac{d}{dx} \left((1-x^2) \frac{d P_l^m(x)}{dx} \right)=
\left(\frac{m}{1-x^2} - l(l+1) \right)P_l^m(x).
\end{equation}
With $x=\cos \theta$, this differential equation results in the nullity of the
expression in Eq. (\ref{condition_bord_1_bis}). This proves that the boundary condition 
$E_\theta(R)=-(r \Omega /c) B_r \sin \theta$ is fulfilled, and that the electromagnetic field derived in section \ref{sec_matching_conditions} is a consistent solution of the problem.

%%%%%%%%%%%%%%%%%%%%%%%%%%%%%%%%%%%%%%%%%%%%%%%%%
%%%%%%%%%%%%%%%%%%%%%%%%%%%%%%%%%%%%%%%%%%%%%%%%%
%%%%%%%%%%%%  REFERENCES   %%%%%%%%%%%%%%%%%%%%%%
%%%%%%%%%%%%%%%%%%%%%%%%%%%%%%%%%%%%%%%%%%%%%%%%%
%%%%%%%%%%%%%%%%%%%%%%%%%%%%%%%%%%%%%%%%%%%%%%%%%

%\end{thebibliography}{}
%\bibliographystyle{alpha} % style aa.bst
%\bibliography{aile_alfven} % your references Yourfile.bib
%\bibliography{article} % your references Yourfile.bib

\end{document}